\documentclass[RNAAS]{aastex631}

\usepackage{amsmath}

\begin{document}

\title{A Dataset for Exploring Stellar Activity in Astrometric Measurements from SDO Images of the Sun}

\author[0009-0007-0022-410X]{Warit Wijitworasart}
\affiliation{Department of Physics and Kavli Institute for Astrophysics and Space Research, Massachusetts Institute of Technology, Cambridge, MA 02139, USA}

\author[0000-0002-7564-6047]{Zoe de Beurs}
\affiliation{Department of Earth, Atmospheric and Planetary Sciences, Massachusetts Institute of Technology, Cambridge, MA 02139, USA}
\affiliation{NSF Graduate Research Fellow and MIT Presidential Fellow}

\author[0000-0001-7246-5438]{Andrew Vanderburg}
\affiliation{Department of Physics and Kavli Institute for Astrophysics and Space Research, Massachusetts Institute of Technology, Cambridge, MA 02139, USA}




\begin{abstract}

We present a dataset for investigating the impact of stellar activity on astrometric measurements using NASA's Solar Dynamics Observatory (SDO) images of the Sun. The sensitivity of astrometry for detecting exoplanets is limited by stellar activity (e.g. starspots), which causes the measured ``center of flux" of the star to deviate from the true, geometric, center, producing false positive detections. We analyze Helioseismic and Magnetic Imager continuum image data obtained from SDO between July 2015 and December 2022 to examine this ``astrometric jitter" phenomenon for the Sun. We employ data processing procedures to clean the images and compute the time series of the sunspot-induced shift between the center of flux and the geometric center. The resulting time series show quasiperiodic variations up to 0.05\% of the Sun's radius at its rotation period.


\end{abstract}




\section{Introduction} \label{sec:intro}

Astrometric planet detection relies on measuring the location of stars in the sky and searching for wobbles due to the gravitational pull of a planet. The ongoing \textit{Gaia} mission is expected to detect thousands of exoplanets using astrometry \citep{perryman, Yahalomi2023}, and several proposed missions hope to launch in the coming decades \citep{theia, ches}. Due to high precision requirements, small perturbations in the measurement may lead to false-positive detections, such as those caused by stellar activity in the form of starspots. Starspots cause the star's flux-based measured position to shift from its geometric center, and will also move with the star's surface as it rotates, creating a quasi-periodic pseudo-wobble motion \citep[e.g.][]{eriksson, morris, Shapiro_2021, Sowmya2021, Kaplan-Lipkin2022, sowmya} called ``astrometric jitter''. We measure this phenomenon for the Sun using the HMI continuum image data from NASA’s SDO satellite, which covers a small range of wavelengths near the 6173\AA\ Fe I line. We computed the time series of the shift between the Sun's center of flux and its geometric center due to sunspots in data from July 2015 to December 2022.

\section{Methods}

To measure the Sun's center of flux and geometric center from SDO images, we took the following steps: 

\begin{itemize}
    \item We downloaded the HMI Continuum images of the Sun, taken at 12:00 AM UTC daily from 1 July 2015 to 31 December 2022, in \verb|fits| files from the SDO archive using \verb|sunpy.net.Fido| \citep{sunpy_community2020}.

    \item To compute the geometric center of the Sun, we modeled the solar images by calculating the radial intensity profile, $I(r)$, of the Sun's disk as a function of the distance, $r$, from its geometric center using the following function:
        \begin{equation}
        I(r) = \begin{cases}
                I_0(1 - u_1(1 - \mu) - u_2 \mu \log\mu) + k ,& (0 \leq r \leq R) \\
                I(R) - \frac{r - R}{s} (I(R) - k),& (R + 0.2 \leq r \leq R + 0.2 + s) \\
                k + \frac{mR}{r} - \frac{mr}{R + 0.2 + s},& (r > R + 0.2 + s)
                \end{cases}
        \end{equation}
        where $I_0$ represents the central intensity, $R$ is the disk radius, $u_1$ and $u_2$ are linear and quadratic limb darkening coefficients, $\mu \equiv \sqrt{1 - \frac{r^2}{R^2}}$, $k$ is the background intensity, $s$ is a ``smear'' coefficient, and $m$ is a ``slope'' coefficient. All length parameters are in pixels. The function was initially computed in steps of 1 pixel in the first part, $s$ pixels in the second part, and 10 pixels in the third part.
        
        We then calculated $r$ of each pixel and used \verb|scipy.interpolate.interp1d| \citep{2020SciPy} to interpolate the function over these distances. This produced a simulated image of the Sun with an intensity profile and center location defined by inputs to our function.
        
        We performed a fit to this intensity profile, the Sun's radius, and the Sun's geometric center coordinates in each image, using \verb|mpfit| (\citealt{mpfit}\footnote{\url{https://github.com/segasai/astrolibpy/blob/master/mpfit/mpfit.py}}). First, we fitted the function to an entire image, minimizing the residual $I_{\text{image}} - I_{\text{model}}$. From this fit, we measured $u_1 = 0.74$ and $u_2 = 0.34$. Then, for every other image, we minimized the residual over the range $0.95R \leq r \leq 1.05R$, holding limb darkening parameters fixed at those values, but varying other parameters (center coordinates $x_{\text{fit}}$ and $y_{\text{fit}}$, $R$, $I_0$, $k$, $s$, and $m$). We evaluated each fit by visual inspection. 

    \item Next, we measured the ``center of flux" -- a good proxy for the location measured by telescopes like \textit{Gaia}. However, we needed to perform additional steps to ``clean" the images of artifacts as follows:
    \begin{enumerate}
        \item We measured a radial intensity profile for one image.
        \item For each other image, we divided the values in each pixel by that intensity profile to flatten the limb-darkening effect of the Sun.
        \item We subtracted the maximum intensity from the image, bringing the disk's brightness to 0. The intensities of bright and dark spots then became slightly above and below 0, respectively.
        \item We clipped out the data beyond 0.999$R_{\text{sun}}$ and ignored it for the rest of the analysis.
        \item On the flattened and continuum-subtracted image, we took the median value of each vertical column and subtracted it from that column. We repeated the step for each horizontal row.
        \item Finally, we added the continuum back and unflattened the image.
    \end{enumerate}
    
    \item Finally, we computed the ``center of flux" coordinates, $x_{\text{cen}}$ and $y_{\text{cen}}$ of each cleaned image using \verb|photutils.centroid.centroid_com| \citep{larry_bradley_2022_6825092}. Our processing removes most of the unwanted varying baseline intensity. To remove the last residual systematics, we high-pass-filtered the time series by fitting and subtracting a basis spline \citep{vj14}.  We then subtracted the geometric center from the high-pass-filtered ``center of flux'' to achieve the time series.
    \end{itemize}
    
\section{Results}
    
\begin{figure*}
    \centering
    \includegraphics[width = 0.9\linewidth]{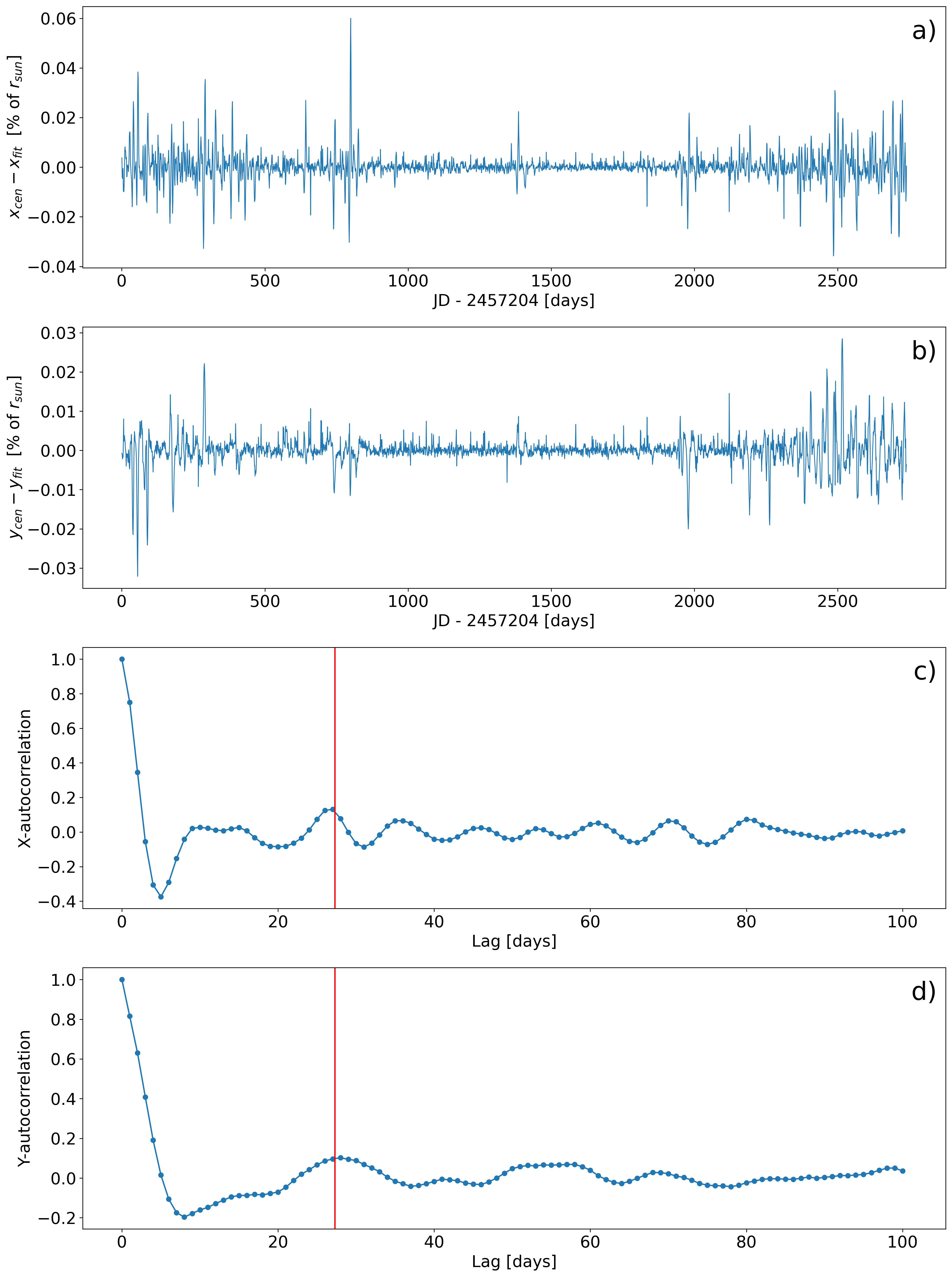}
    \caption{a) Time series of $x_{\text{cen}} - x_{\text{fit}}$ as a percentage of $r_{\text{sun}}$ b) Time series of $y_{\text{cen}} - y_{\text{fit}}$ as a percentage of $r_{\text{sun}}$ c) Autocorrelation value of $x_{\text{cen}} - x_{\text{fit}}$ vs Time Lag d) Autocorrelation value of $y_{\text{cen}} - y_{\text{fit}}$ vs Time Lag. Red vertical lines in c) and d) indicate the Sun's Carrington rotation period of 27.2753 days.}
    \label{fig:my_label}
\end{figure*}

Our results are shown in Figure \ref{fig:my_label}. We plotted the time series of the x and y-coordinates shift ($x_{\text{cen}} - x_{\text{fit}}$ and $y_{\text{cen}} - y_{\text{fit}}$) as a percentage of the Sun's radius $r_{\text{sun}}$ as well as the autocorrelation periodograms using \verb|statsmodels.graphics.tsaplots.plot_acf| (\citealt{statsmodels}). The time series show ``astrometric jitter" up to 0.05\% of the Sun's radius, and the autocorrelation functions both show peaks at the Sun's rotation period.

\section{acknowledgments}
We acknowledge support from MIT/UROP, the MIT Presidential Fellowship, and the NSF GRFP (No 1745302) and make use of SDO data, NASA's Astrophysics Data System, Photutils, and SunPy.

%





\bibliography{main}{}
\bibliographystyle{aasjournal}



\end{document}